\documentclass[twocolumn,showpacs,preprintnumbers,pra,hyperref]{revtex4}
\usepackage{amssymb}
\usepackage{amsfonts}
\usepackage{amsmath}
\usepackage{graphicx}

\begin{document}

\title{Light field distribution of general function photonic crystals}
\author{Xiang-Yao Wu$^{a}$\thanks{%
E-mail: wuxy2066@163.com}, Bo-Jun Zhang$^{a}$, Xiao-Jing
Liu$^{a}$\\ Si-Qi Zhang$^{a}$, Jing Wang$^{a}$, Nuo Ba$^{a}$, Li
Xiao$^{a}$ and Hong Li$^{a}$} \affiliation{$^{a}${\small Institute
of Physics, Jilin Normal University, Siping 136000, China}
 }

\begin{abstract}
In this paper, We have presented a new general function photonic
crystals (GFPCs), which refractive indexes are line functions of
space position in two mediums $A$ and $B$, and obtain new results:
(1) when the line function of refractive indexes is up or down,
the transmissivity can be far larger or smaller than $1$. (2) when
the refractive indexes function increase or decrease along the
direction of incident light, the light intensity should be
magnified or weaken, which can be made optical magnifier or
attenuator. (3) The GFPCs can be made optical diode when the light
positive and negative incident the GFPCs.\\

\end{abstract}

\maketitle

\maketitle {\bf 1. Introduction} \vskip 8pt

Photonic crystals (PC) are a new kind of materials which
facilitate the control of the light [1-5]. An important feature of
the photonic crystals is that there are allowed and forbidden
ranges of frequencies at which light propagates in the direction
of index periodicity [6-9]. Due to the forbidden frequency range,
known as photonic band gap (PBG) [10-15], which forbids the
radiation propagation in a specific range of frequencies. The
existence of PBGs will lead to many interesting phenomena. In the
past ten years has been developed an intensive effort to study and
micro-fabricate PBG materials in one, two or three dimensions,
e.g., modification of spontaneous emission [16-19] and photon
localization [20-23]. Reduction or suppression of the density of
states within the band gap facilitates light localization and
trapping in a bulk material [24-25], as well as the inhibition of
spontaneous emission over a broad frequency range. Thus numerous
applications of photonic crystals have been proposed in improving
the performance of optoelectronic and microwave devices such as
high-efficiency semiconductor lasers, right emitting diodes, wave
guides, optical filters, high-Q resonators, antennas,
frequency-selective surface, optical limiters and amplifiers
[26-27].

In Ref. [28], we have proposed special function photonic crystals,
which the medium refractive index is the function of space
position, but the function value of refractive index is equal at
two endpoints of every medium $A$ and $B$. In Ref. [29], we have
proposed a new general function photonic crystals, i.e., the
medium refractive index is a arbitrary function of space position,
and find the transmissivity of one-dimensional GFPCs can be much
larger or smaller than $1$ for different slope linearity
refractive index function. In this paper, we choose two linearity
refractive index functions for two  medium $A$ and $B$, and give
the light field distribution in the GFPCs. We obtain some new
results: (1) when the line function of refractive indexes is up,
the transmissivity can be far larger than $1$. (2) when the line
function of refractive indexes is down, the transmissivity can be
far smaller than $1$. (3) when the two-endpoint values of
refractive index are equal for mediums $A$ and $B$, the
transmissivity $T$ is in the range of $0$ and $1$ (it becomes
conventional PCs ). So, the conventional PCs is the special case
of the GFPCs. (4) when the refractive indexes function increase
along the direction of incident light, the light intensity should
be magnified, which can be made light magnifier. (5) when the
refractive indexes function decrease along the direction of
incident light, the light intensity should be weaken, which can be
made light attenuator. (5) The GFPCs can be made light diode. The
general function photonic crystals can be applied to design more
optical instruments.

\vskip 8pt

{\bf 2. The light motion equation in general function photonic
crystals} \vskip 8pt

For the general function photonic crystals, the medium refractive
index is a periodic function of the space position, which can be
written as $n(z)$, $n(x, z)$ and $n(x, y, z)$ corresponding to
one-dimensional, two-dimensional and three-dimensional function
photonic crystals. In the following, we shall deduce the light
motion equations of the one-dimensional general function photonic
crystals, i.e., the refractive index function is $n=n(z)$,
meanwhile motion path is on $xz$ plane. The incident light wave
strikes plane interface point $A$, the curves $AB$ and $BC$ are
the path of incident and reflected light respectively, and they
are shown in FIG. 1.
\begin{figure}[tbp]
\includegraphics[width=8.5 cm]{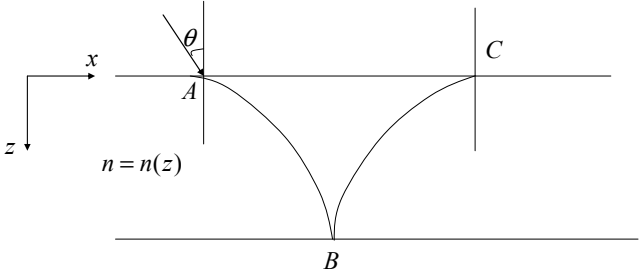}
\caption{The motion path of light in the medium of refractive
index $n(z)$.}
\end{figure}

The light motion equation can be obtained by Fermat principle, it
is
\begin{eqnarray}
\delta\int^{B}_{A}n(z) ds=0.\label{1}
\end{eqnarray}
In the two-dimensional transmission space, the line element $ds$
is
\begin{eqnarray}
ds=\sqrt{(dx)^{2}+(dz)^{2}}=\sqrt{1+\dot{z}^{2}}dx,
\end{eqnarray}
where $\dot{z}=\frac{dz}{dx}$, then Eq. (1) becomes
\begin{eqnarray}
\delta\int^{B}_{A}n(z)\sqrt{1+(\dot{z})^{2}}dx=0.
\end{eqnarray}
The Eq. (3) change into
\begin{eqnarray}
\int^{B}_{A}(\frac{\partial(n(z)\sqrt{1+\dot{z}^{2}})}{\partial
z}\delta z+\frac{\partial(n(z)\sqrt{1+\dot{z}^{2}})}{\partial
\dot{z}}\delta\dot{z})dx=0,
\end{eqnarray}
At the two end points $A$ and $B$, their variation is zero, i.e.,
$\delta z (A)=\delta z (B)=0$. For arbitrary variation $\delta z$, the Eq. (4) becomes \\
\begin{eqnarray}
&&\frac{dn(z)}{dz}\sqrt{1+\dot{z}^{2}} -\frac{d n(z)}{d
z}\dot{z}^{2}(1+\dot{z}^{2})^{-\frac{1}{2}}
\nonumber\\&&-n(z)\frac{\ddot{z}\sqrt{1+\dot{z}^{2}}
-\dot{z}^{2}\ddot{z}(1+\dot{z}^{2})^{-\frac{1}{2}}}{1+\dot{z}^{2}}
=0,
\end{eqnarray}
simplify Eq. (5), we have
\begin{eqnarray}
\frac{d n(z)}{n(z)} = \frac{\dot{z}d\dot{z}}{1+\dot{z}^{2}}.
\end{eqnarray}\\
The Eq. (6) is light motion equation in one-dimensional function
photonic crystals.
\begin{figure}[tbp]
\includegraphics[width=8.5 cm]{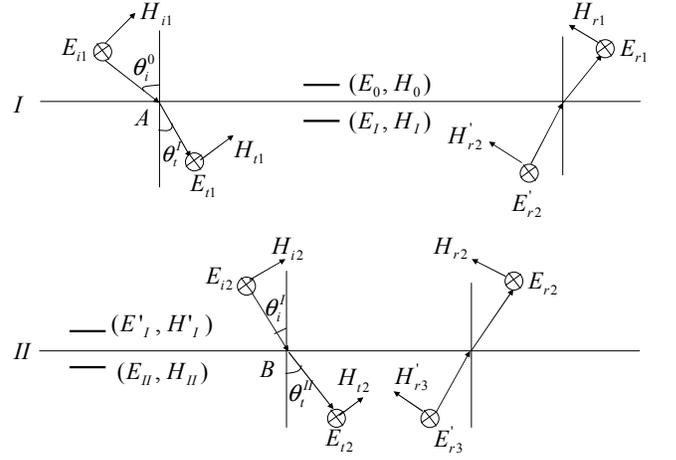}
\caption{The light transmission and electric magnetic field
distribution figure in FIG.1 medium.} \label{Fig1}
\end{figure}
\vskip 8pt

{\bf 3. The transfer matrix of one-dimensional general function
photonic crystals} \vskip 8pt

In this section, we should calculate the transfer matrix of
one-dimensional general function photonic crystals. In fact, there
is the reflection and refraction of light at a plane surface of
two media with different dielectric properties. The dynamic
properties of the electric field and magnetic field are contained
in the boundary conditions: normal components of $D$ and $B$ are
continuous; tangential components of $E$ and $H$ are continuous.
We consider the electric field perpendicular to the plane of
incidence, and the coordinate system and symbols as shown in FIG.
2.

On the two sides of interface I, the tangential components of
electric field $E$ and magnetic field $H$ are continuous, there
are

\begin{eqnarray}
\left \{ \begin{array}{ll}
 E_{0}=E_{I}=E_{t1}+E'_{r2}\\
H_{0}=H_{I}=H_{t1}\cos\theta_{t}^{I}-H'_{r2}\cos\theta_{t}^{I}.
\end{array}
\right.
\end{eqnarray}
On the two sides of interface II, the tangential components of
electric field $E$ and magnetic field $H$ are continuous, and give
\begin{eqnarray}
\left \{ \begin{array}{ll}
 E_{II}=E'_{I}=E_{i2}+E_{r2}\\
H_{II}=H'_{I}=H_{i2}\cos\theta_{i}^{I}-H_{r2}\cos\theta_{i}^{I},
\end{array}
\right.
\end{eqnarray}
the electric field ${E_{t1}}$ is
\begin{eqnarray}
E_{t1}=E_{t10}{e^{i(k_{x}x_{A}+k_{z}z)}|_{z=0}}=E_{t10}e^{i\frac{\omega}{c}n(0)\sin\theta_{t}^{I}x_{A}},
\end{eqnarray}
and the electric field ${E_{i2}}$ is
\begin{eqnarray}
E_{i2}&=&E_{t10}{e^{i(k'_{x}x_{B}+k'_{z} z)}|_{z=b}}
\nonumber\\&=&E_{t10}e^{i\frac{\omega}{c}n(b)(\sin\theta_{i}^{I}x_{B}+\cos\theta_{i}^{I}
b)}.
\end{eqnarray}
Where $x_{A}$ and $x_{B}$ are $x$ component coordinates
corresponding to point $A$ and point $B$. We should give the
relation between $E_{i2}$ and $E_{t1}$. By integrating the two
sides of Eq. (6), we can obtain the coordinate component $x_{B}$
of point $B$
\begin{eqnarray}
\int^{n(z)}_{n(0)}\frac{dn(z)}{n(z)}=\int^{k_{z}}_{k_{0}}\frac{\dot{z}d\dot{z}}{1+\dot{z}^{2}},
\end{eqnarray}
to get
\begin{eqnarray}
k_z^2=(1+k_0^2)(\frac{n(z)}{n(0)})^2-1,
\end{eqnarray}
and
\begin{eqnarray}
dx=\frac{dz}{\sqrt{(1+k_{0}^{2})(\frac{n(z)}{n(0)})^{2}-1}}.
\end{eqnarray}
where $k_{0}=\cot\theta_{t}^{I}$ and $k_{z}=\frac{d z}{d x}$ From
Eq. (12), there is $n(z)>n(0)\sin\theta^{I}_{t}$. and the
coordinate $x_{B}$ is
\begin{eqnarray}
x_{B}=x_{A}+\int^{b}_{0}\frac{dz}{\sqrt{(1+k_{0}^{2})(\frac{n(z)}{n(0)})^{2}-1}},
\end{eqnarray}
where $b$ is the medium thickness of FIG. 1 and FIG. 2.\\
By substituting Eqs. (9) and (14)into (10), and using the equality
\begin{eqnarray}
 n(0)\sin\theta_{t}^{I}=n(b)\sin\theta_{i}^{I},
\end{eqnarray}
  we have
\begin{eqnarray}
E_{i2}&=&E_{t1}e^{i{\delta}_{b}},
\end{eqnarray}
where
\begin{figure}[tbp]
\includegraphics[width=8 cm]{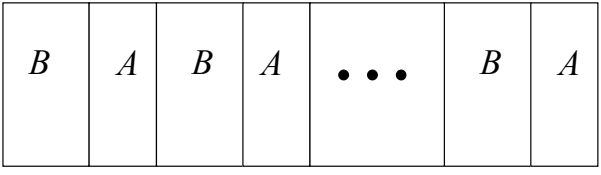}
\caption{The structure $(BA)^{N}$ of the general function photonic
crystals.}
\end{figure}
\begin{eqnarray}
\delta_{b}=\frac{\omega}{c}n_{b}(b)(\cos\theta_{i}^{I}b+\sin\theta_{i}^{I}
\int^{b}_{0}\frac{dz}{\sqrt{\frac{n_{b}^{2}(z)}{n_{0}^{2}\sin^{2}\theta_{i}^{0}}-1}}),
\end{eqnarray}
and similarly
\begin{eqnarray}
E'_{r2}=E_{r2}e^{i\delta_{b}}.
\end{eqnarray}
Substituting Eqs. (16) and (18) into (7) and (8), and using
$H=\sqrt{\frac{\varepsilon_{0}}{\mu_{0}}}nE$, we obtain
\begin{eqnarray}
\left(%
\begin{array}{c}
  E_{I} \\
  H_{I} \\
\end{array}%
\right)&=&M_{B}\left(%
\begin{array}{c}
  E_{II} \\
  H_{II} \\
\end{array}%
\right),
\end{eqnarray}
where
\begin{eqnarray}
M_{B}=\left(%
\begin{array}{cc}
 \cos\delta_{b} & -\frac{i\sin\delta_{b}}{\sqrt{\frac{\varepsilon_{0}}{\mu_{0}}}n_{b}(b)\cos\theta_{i}^{I}} \\
 -in_{b}(0)\sqrt{\frac{\varepsilon_{0}}{\mu_{o}}}\cos\theta_{t}^{I}\sin\delta_{b}
 & \frac{n_{b}(0)\cos\theta_{t}^{I}\cos\delta_{b}}{n_{b}(b)\cos\theta_{i}^{I}}\\
\end{array}%
\right),
\end{eqnarray}
The Eq. (20) is the transfer matrix $M$ in the medium of FIG. 1
and FIG. 2. By refraction law, we can obtain
\begin{eqnarray}
\sin\theta^{I}_{t}=\frac{n_{0}}{n(0)}\sin\theta^{0}_{i},\cos\theta^{I}_{t}
=\sqrt{1-\frac{n_{0}^{2}}{n^{2}(0)}\sin^{2}\theta^{I}_{t}},
\end{eqnarray}
where $n_0$ is air refractive index, and $n(0)=n(z)|_{z=0}$. Using
Eqs. (15) and (21), we can calculate $\cos\theta_{i}^{I}$.

\begin{figure}[tbp]
\includegraphics[width=8 cm]{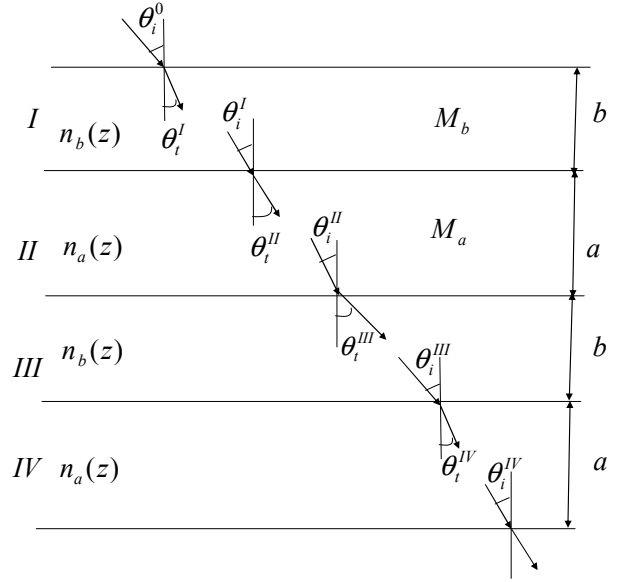}
\caption{The two periods transmission figure of light in general
function photonic crystals.}
\end{figure}
\vskip 8pt

{\bf 4. The transmissivity and light field distribution of
one-dimensional general function photonic crystals} \vskip 8pt

In section 3, we obtain the $M$ matrix of the half period. We know
that the conventional photonic crystals is constituted by two
different refractive index medium, and the refractive indexes are
not continuous on the interface of the two mediums. We could
devise the one-dimensional general function photonic crystals
structure as follows: in the first half period, the refractive
index distributing function of medium $B$ is $n_{b}(z)$. and in
the second half period, the refractive index distributing function
of medium $A$ is $n_{a}(z)$, corresponding thicknesses are $b$ and
$a$, respectively. Their refractive indexes satisfy condition
$n_{b}(b)\neq n_{a}(0)$, their structure are shown in FIG. 3, and
FIG. 4. The Eq. (20) is the half period transfer matrix of medium
$B$. Obviously, the half period transfer matrix of medium A is
\begin{eqnarray}
M_{A}=\left(%
\begin{array}{cc}
 \cos\delta_{a} & -\frac{i\sin\delta_{a}}{\sqrt{\frac{\varepsilon_{0}}{\mu_{0}}}n_{a}(a)\cos\theta_{i}^{II}} \\
 -in_{a}(0)\sqrt{\frac{\varepsilon_{0}}{\mu_{o}}}\cos\theta_{t}^{II}\sin\delta_{a}
 & \frac{n_{a}(0)\cos\theta_{t}^{II}\cos\delta_{a}}{n_{a}(a)\cos\theta_{i}^{II}}\\
\end{array}%
\right),
\end{eqnarray}
where
\begin{eqnarray}
\delta_{a}&=&\frac{\omega}{c}n_{a}(a)[\cos\theta^{II}_{i}\cdot a
\nonumber\\&&
+\sin\theta^{II}_{i}\int^{a}_{0}\frac{dz}{\sqrt{\frac{n_{a}^{2}(z)}{n_{0}^{2}\sin^{2}\theta_{i}^{0}}-1}}],
\end{eqnarray}
\begin{eqnarray}
\cos\theta^{II}_{t}
=\sqrt{1-\frac{n_{0}^{2}}{n_{a}^{2}(0)}\sin^{2}\theta_{i}^{0}},
\end{eqnarray}
and
\begin{eqnarray}
\sin\theta^{II}_{i}=\frac{n_{0}}{n_{a}(a)}\sin\theta_{i}^{0},
\end{eqnarray}
\begin{eqnarray}
\cos\theta^{II}_{i}
=\sqrt{1-\frac{n_{0}^{2}}{n_{a}^{2}(a)}\sin^{2}\theta_{i}^{0}}.
\end{eqnarray}
In one period, the transfer matrix $M$ is
\begin{eqnarray}
&&M=M_{B}\cdot M_{A}\nonumber\\
&&=\left(%
\begin{array}{cc}
  \cos\delta_{b} & \frac{-i\sin\delta_{b}}{\sqrt{\frac{\varepsilon_{0}}{\mu_{0}}}n_{b}(b)\cos\theta_{i}^{I}} \\
 -in_{b}(0)\sqrt{\frac{\varepsilon_{0}}{\mu_{o}}}\cos\theta_{t}^{I}\sin\delta_{b}
 & \frac{n_{b}(0)\cos\theta_{t}^{I}\cos\delta_{b}}{n_{b}(b)\cos\theta_{i}^{I}}\\
\end{array}%
\right) \nonumber\\&&
\left(%
\begin{array}{cc}
   \cos\delta_{a} & \frac{-i\sin\delta_{a}}{\sqrt{\frac{\varepsilon_{0}}{\mu_{0}}}n_{a}(a)\cos\theta_{i}^{II}} \\
 -in_{a}(0)\sqrt{\frac{\varepsilon_{0}}{\mu_{o}}}\cos\theta_{t}^{II}\sin\delta_{a}
 & \frac{n_{a}(0)\cos\theta_{t}^{II}\cos\delta_{a}}{n_{a}(a)\cos\theta_{i}^{II}}\\
\end{array}%
\right).
\end{eqnarray}
The form of the GFPCs transfer matrix $M$ is more complex than the
conventional PCs. The angle $\theta_{t}^{I}$, $\theta_{i}^{I}$,
$\theta_{t}^{II}$ and $\theta_{i}^{II}$ are shown in Fig. 4. The
characteristic equation of GFPCs is
\begin{eqnarray}
\left(%
\begin{array}{c}
  E_{1} \\
  H_{1} \\
\end{array}%
\right)&=&M_{1}M_{2}\cdot\cdot\cdot M_{N}
\left(%
\begin{array}{c}
  E_{tN+1} \\
  H_{tN+1} \\
\end{array}%
\right) \nonumber\\&=&M_{b}M_{a}M_{b}M_{a}\cdot\cdot\cdot M_{b}M_{a}\left(%
\begin{array}{c}
  E_{tN+1} \\
  H_{tN+1} \\
\end{array}%
\right)
\nonumber\\&=&M\left(%
\begin{array}{c}
  E_{tN+1} \\
  H_{tN+1} \\
\end{array}%
\right)=\left(%
\begin{array}{c c}
  A &  B \\
 C &  D \\
\end{array}%
\right)
 \left(%
\begin{array}{c}
  E_{tN+1} \\
  H_{tN+1} \\
\end{array}%
\right).
\end{eqnarray}
Where $N$ is the period number.
\begin{figure}[tbp]
\includegraphics[width=9 cm]{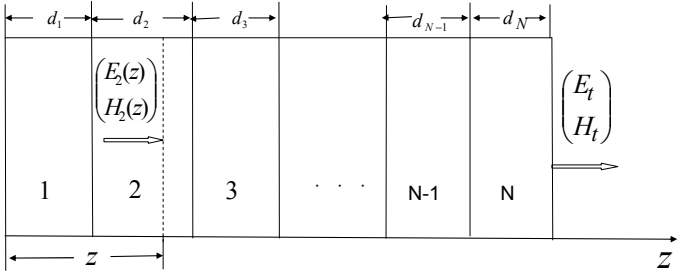}
\caption{The input and output light in the GFPCs.}
\end{figure}
\begin{figure}[tbp]
\includegraphics[width=9 cm]{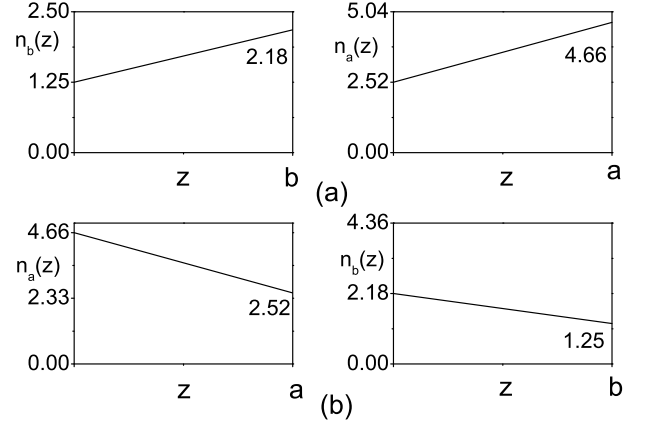}
\caption{The line refractive index functions in a period. The FIG
.6(a) is the up line function of refractive indexes, and FIG .6(b)
is the down line function of refractive indexes.}
\end{figure}
With the transfer matrix $M$ (Eq. (28)), we can obtain the
transmission and reflection coefficient $t$ and $t$, and the
transmissivity and reflectivity $T$ and $R$, they are
\begin{eqnarray}
t=\frac{E_{tN+1}}{E_{0i}}=\frac{2\eta_{0}}{A\eta_{0}+B\eta_{0}\eta_{N+1}+C+D\eta_{N+1}},
\end{eqnarray}
\begin{eqnarray}
r=\frac{E_{0r}}{E_{0i}}=\frac{A\eta_{0}+B\eta_{0}\eta_{N+1}-C-D\eta_{0}}{A\eta_{0}+B\eta_{0}\eta_{N+1}+C+D\eta_{0}},
\end{eqnarray}
and
\begin{eqnarray}
T=t\cdot t^{*},
\end{eqnarray}
\begin{eqnarray}
R=r\cdot r^{*}.
\end{eqnarray}
Where $\eta_{0}=\eta_{N+1}=\sqrt{\frac{\varepsilon_0}{\mu_0}}$,
$E_{0i}$ and $E_{0r}$ are incident and reflection electric field,
and $E_1=E_{0i}+E_{0r}$. In the following, we give the electric
field distribution of light in the one-dimensional GFPCs, and the
propagation figure of light is shown in FIG. 5. From Eq. (28), we
have
\begin{eqnarray}
&&\left(%
\begin{array}{c}
  E_{2} (z)\\
  H_{2} (z)\\
\end{array}%
\right)=M_{1}(d_{1})M_{2}(d_{2}-z)\cdot\cdot\cdot M_{N}(d_{N})
\left(%
\begin{array}{c}
  E_{tN+1} \\
  H_{tN+1} \\
\end{array}%
\right)
 \nonumber\\&=&\left(%
\begin{array}{c c}
  A(z) &  B(z) \\
 C(z) &  D(z) \\
\end{array}%
\right)
 \left(%
\begin{array}{c}
  E_{tN+1} \\
  H_{tN+1} \\
\end{array}%
\right).
\end{eqnarray}
where $d_1$ and $d_2$ are the thickness of first and second
medium, respectively, $E_2(z)$ and $H_2(z)$ are the electric field
and magnetic field in the second medium, when position $z$ changes
we can obtain the electric field and magnetic field in other
medium. The $E_t$ and $H_t$ are the transmission electric and
magnetic field. Eq. (33) can be written as
\begin{eqnarray}
E(z)&=&A(z)E_{tN+1}+B(z)H_{tN+1}\nonumber\\&=&
A(z)E_{tN+1}+B(z)\sqrt{\frac{\varepsilon_{0}}{\mu_{0}}}E_{tN+1}
\nonumber\\&=&(A(z)+B(z)\sqrt{\frac{\varepsilon_{0}}{\mu_{0}}})tE_{0i},
\end{eqnarray}
and then
\begin{eqnarray}
|\frac{E(z)}{E_{0i}}|^{2}=|t|^{2}|A(z)+B(z)\sqrt{\frac{\varepsilon_{0}}{\mu_{0}}})|^{2}
\end{eqnarray}

{\bf 5. Numerical result}

\vskip 8pt

In this section, we report  our numerical results of
transmissivity. We consider refractive indexes of the linearity
functions in a period, it is

\begin{eqnarray}
n_{b}(z)=n_{b}(0)+\frac{n_{b}(b)-n_{b}(0)}{b}z, \hspace{0.1in} 0
\leq z\leq b,
\end{eqnarray}
\begin{eqnarray}
n_{a}(z)=n_{a}(0)+\frac{n_{a}(a)-n_{a}(0)}{a}z, \hspace{0.1in} 0
\leq z\leq a,
\end{eqnarray}
Eqs. (36) and (37) are the line refractive indexes distribution
functions of two half period mediums $B$ and $A$. When the
endpoint values $n_{b}(0)$, $n_{b}(b)$, $n_{a}(0)$ and $n_{a}(a)$
are all given, the line refractive index functions $n_{b}(z)$ and
$n_{a}(z)$ are ascertained. The main parameters are: the half
period thickness $b$ and $a$, the starting point refractive
indexes $n_{b}(0)$ and $n_{a}(0)$, and end point refractive
indexes $n_{b}(b)$ and $n_{a}(a)$, the optical thickness of the
two mediums are equal, i.e., $n_{b}(0)b=n_{a}(0)a$, the incident
angle $\theta_{i}^{0}=0$, the center frequency
$\omega_{0}=1.215\times10^{15} Hz$, the thickness $b=310 nm$,
$a=154 nm$ and the period number $N=16$, i.e., the structure of
GFPCs is $(BA)^{16}$.

\begin{figure}[tbp]
\includegraphics[width=9 cm]{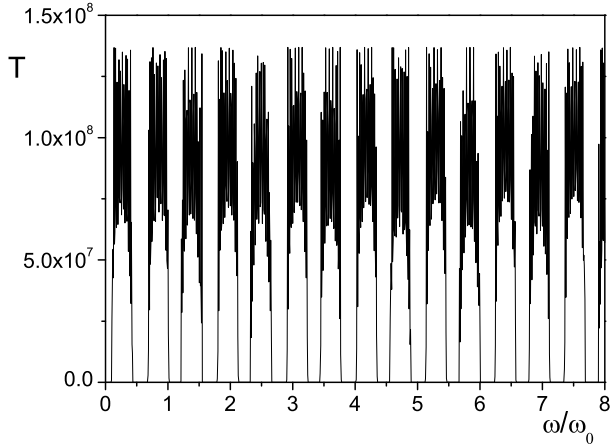}
\caption{The relation between transmissivity and frequency
corresponding to the up line function of refractive indexes (FIG
.6(a)). }
\end{figure}

\begin{figure}[tbp]
\includegraphics[width=9 cm]{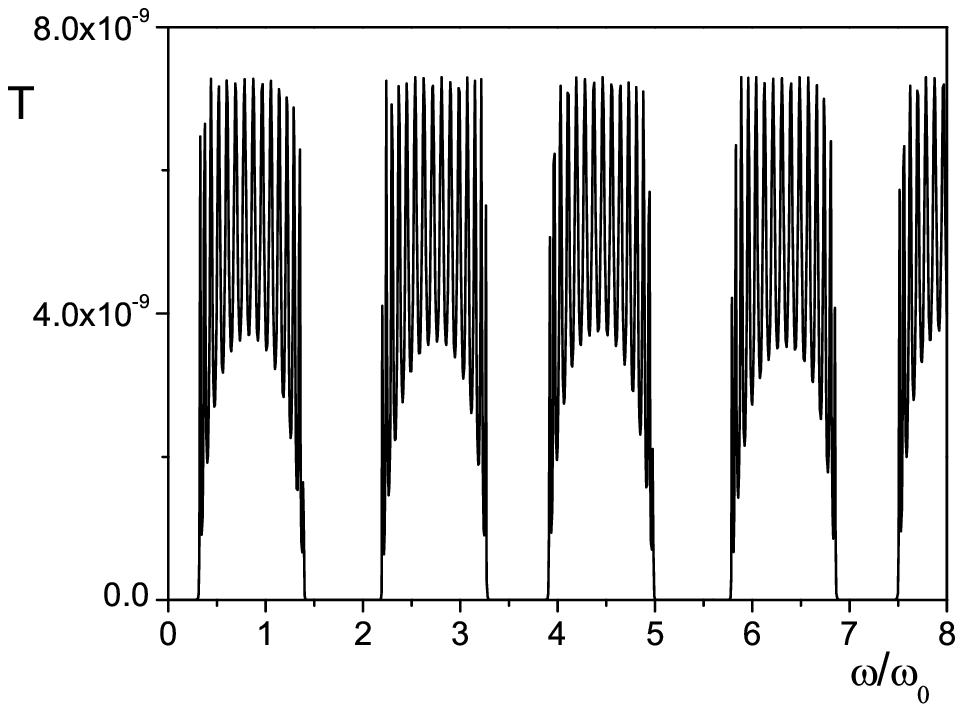}
\caption{The relation between transmissivity and frequency
corresponding to the down line function of refractive indexes (FIG
.6(b)). }
\end{figure}

\begin{figure}[tbp]
\includegraphics[width=8.5cm]{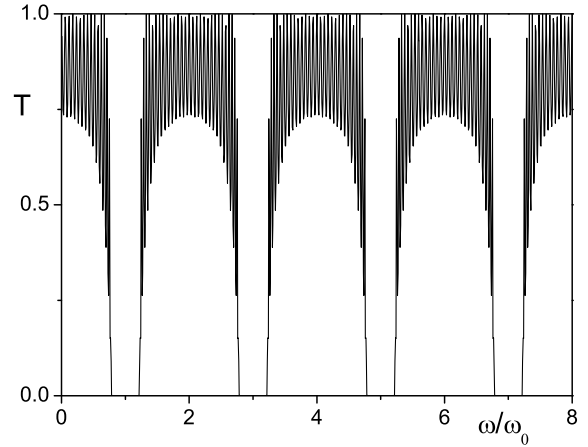}
\caption{The relation between transmissivity and frequency for the
GFPCs, when the two-endpoint values of refractive index are equal
}
\end{figure}

\begin{figure}[tbp]
\includegraphics[width=8.5 cm]{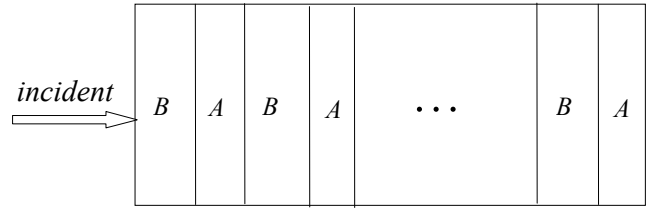}
\caption{The light positive incident to the GFPCs.}
\end{figure}

\begin{figure}[tbp]
\includegraphics[width=8.5 cm]{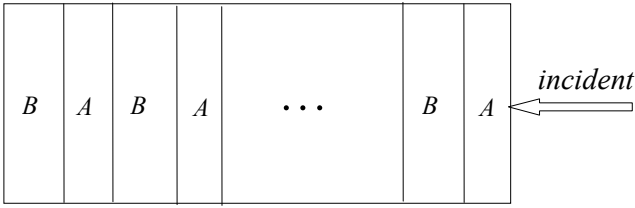}
\caption{The light negative incident to the GFPCs.}
\end{figure}

\begin{figure}[tbp]
\includegraphics[width=8.5 cm]{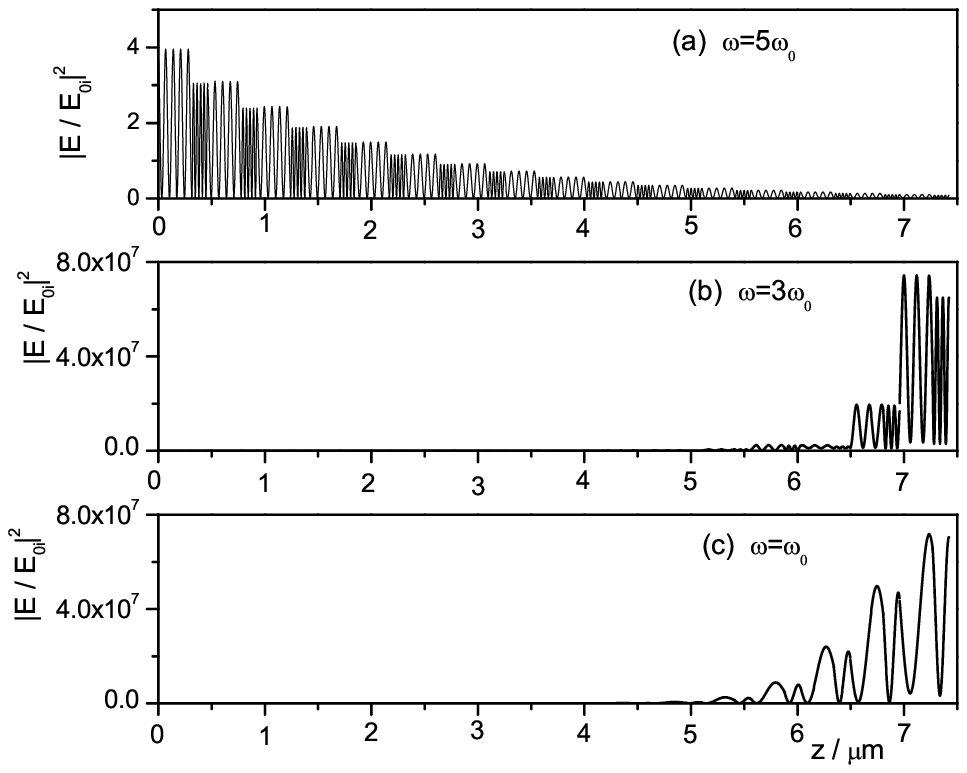}
\caption{The light distribution of positive incident (FIG. 10) in
the GFPCs. Figure (a) (b) and (c) are corresponding to the
incident light frequency $\omega=\omega_{0}$, $3\omega_{0}$ and
$5\omega_{0}$.}
\end{figure}

\begin{figure}[tbp]
\includegraphics[width=8.5 cm]{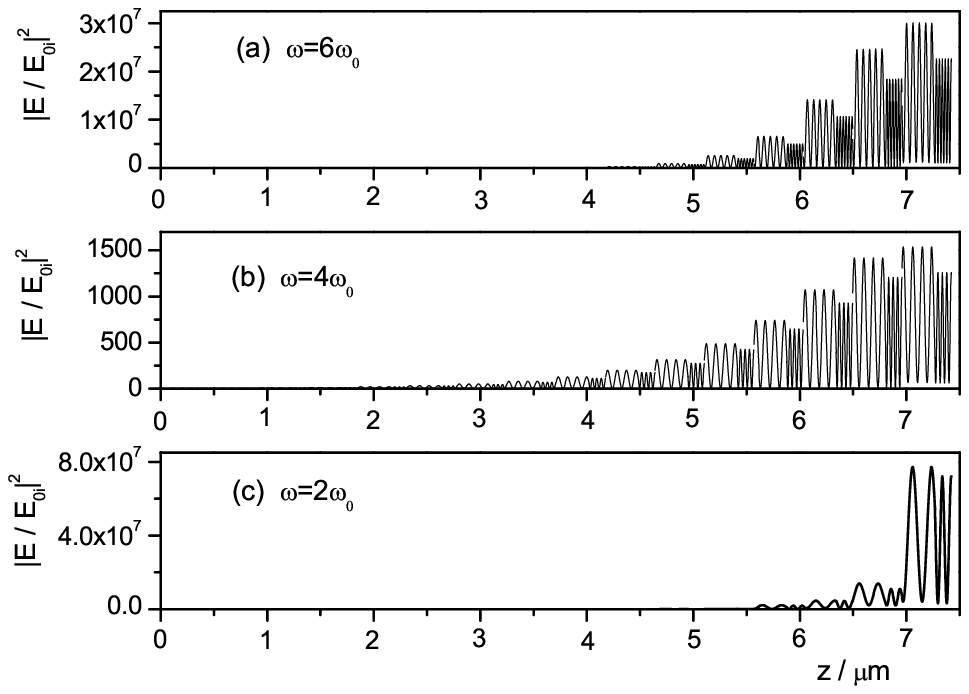}
\caption{The light distribution of positive incident (FIG. 10) in
the GFPCs. Figure (a) (b) and (c) are corresponding to the
incident light frequency $\omega=2\omega_{0}$, $4\omega_{0}$ and
$6\omega_{0}$.}
\end{figure}

\begin{figure}[tbp]
\includegraphics[width=8.5 cm]{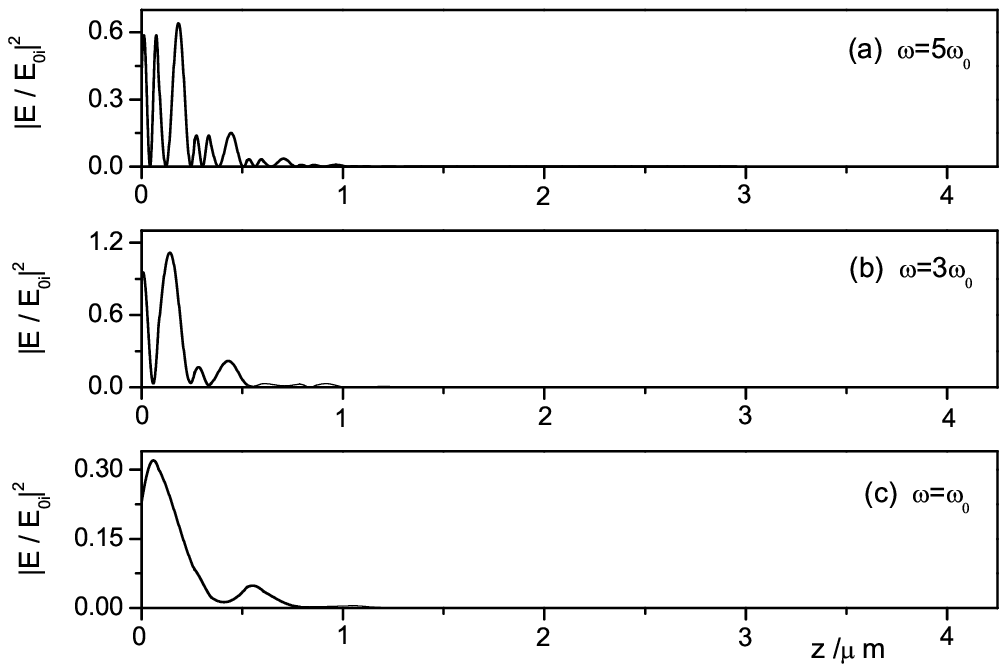}
\caption{The light distribution of negative incident (FIG. 11) in
the GFPCs. Figure (a) (b) and (c) are corresponding to the
incident light frequency $\omega=\omega_{0}$, $3\omega_{0}$ and
$5\omega_{0}$.}
\end{figure}

\begin{figure}[tbp]
\includegraphics[width=8.5 cm]{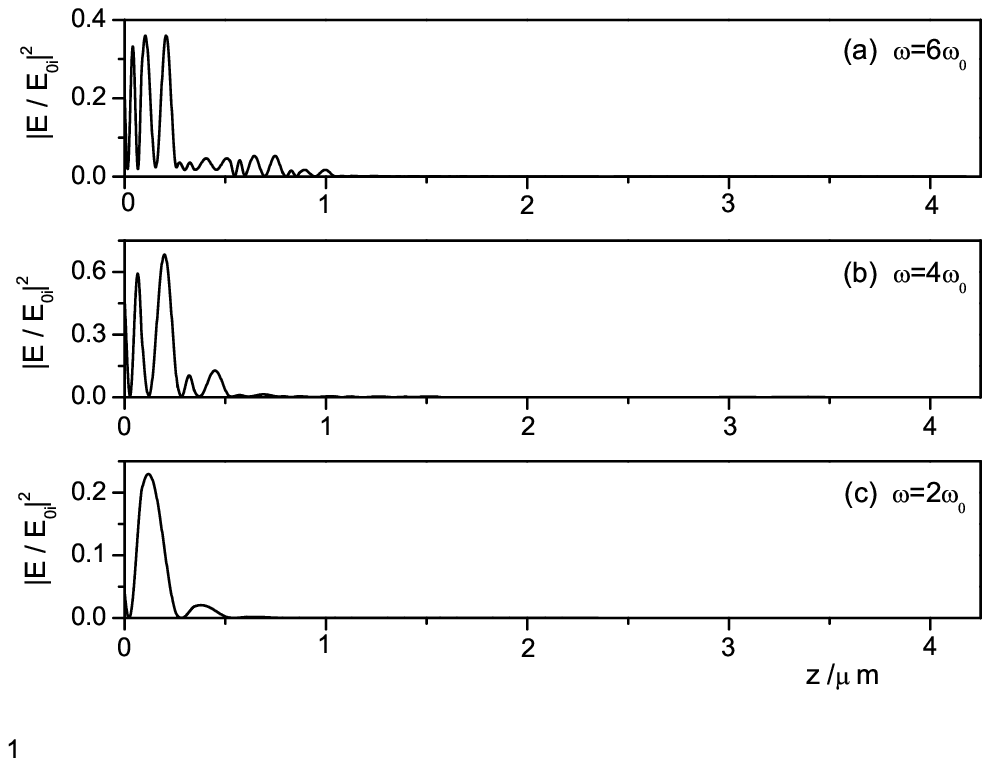}
\caption{The light distribution of negative incident (FIG. 11) in
the GFPCs. Figure (a) (b) and (c) are corresponding to the
incident light frequency $\omega=2\omega_{0}$, $4\omega_{0}$ and
$6\omega_{0}$.}
\end{figure}

In FIG. 5(a), we take $n_{b}(0)=1.25$, $n_{b}(b)=2.18$ for the
medium $B$, and $n_{a}(0)=2.52$, $n_{a}(a)=4.66$ for the medium
$A$, which are the up line function of refractive indexes. In FIG.
5(b), we take $n_{b}(0)=2.18$, $n_{b}(b)=1.25$ for the medium $B$,
and $n_{a}(0)=4.66$, $n_{a}(a)=2.52$ for the medium $A$, which are
the down line function of refractive indexes. By the refractive
indexes function, we can calculate the transmissivity. With the
FIG. 6(a) and 6(b), we obtain the transmissivity distribution in
FIG. (7) and FIG. (8) respectively. From FIG. (7) and FIG. (8), we
obtain the results: (1) when the line function of refractive
indexes is up, the transmissivity can be far larger than $1$ ($T$
maximum is $10^{8}$). (2) when the line function of refractive
indexes is down, the transmissivity can be far smaller than $1$
($T$ maximum is $10^{-9}$), which are different from the
transmissivity of conventional PCs $(0\leq T\leq1)$. These results
are inexplicable and doubtful even, but they are correct. When we
take $n_{b}(0)=n_{b}(b)=1.25$ and $n_{a}(0)=n_{a}(a)=2.52$, i.e.,
two-endpoint values of refractive index are equal for mediums $A$
and $B$ (it is conventional PCs), and by the Eq. (27), (30) and
(32), we obtain the transmissivity distribution in FIG. (9), its
transmissivity $T$ is in the range of $0$ and $1$ (it becomes
conventional PCs transmissivity distribution). So, the
conventional PCs is the special case of the GFPCs.

In the following, we shall study the light field distribution of
the one-dimensional GFPCs for the positive and negative incident
of light. The positive incident figure is shown in FIG. 10, and
the negative incident is shown in FIG. 11, which relative to the
positive incident FIG. 10. The refractive indexes line function of
positive incident is in FIG. 6(a), and the FIG. 6(b) is the
refractive indexes line function of negative incident
corresponding to the positive incident FIG. 6(a). From Eq. (35),
we can calculate the electric field distribution in the GFPCs.
Figures 12 and 13 are the electric field distributions of positive
incident (FIG. 10). In FIG. 12, the frequency of incident light
$\omega$ is the odd times of the center frequency $\omega_{0}$. In
FIG. 12 (a), (b) and (c), the incident light $\omega$ are
corresponding to $\omega_{0}$, $3\omega_{0}$ and $5\omega_{0}$.
When $\omega=\omega_{0}$ and $3\omega_{0}$, the light intensity of
transmission are enhanced or magnified (magnification $M=10^{7}$),
while $\omega=5\omega_{0}$, the intensity of transmission light is
weaken. From FIG. 7, we can find when $\omega=\omega_{0}$ and
$3\omega_{0}$ the light is in conductance band, and the
transmission light has been enhanced by the GFPCs. In the
conventional PCs, the intensity of transmission light can not be
magnified. When $\omega=5\omega_{0}$ the light is in forbidden
band, and the transmission light has been weaken by the GFPCs. In
FIG. 13, the frequency of incident light $\omega$ is the even
times of the center frequency $\omega_{0}$. In FIG. 13 (a), (b)
and (c), the incident light $\omega$ are corresponding to
$2\omega_{0}$, $4\omega_{0}$ and $6\omega_{0}$, the intensities of
transmission light have been magnified. Figures 14 and 15 are the
electric field distributions of negative incident (FIG. 11). In
Figures. 14 and 15, the frequency of incident light $\omega$ is
the odd and even times of the center frequency $\omega_{0}$, and
the intensities of transmission light have been all weaken. By
calculation, we can obtain the following results for the GFPCs:
(1) For the positive incident, i.e., the refractive indexes
increasing along the direction of incident light, the intensity of
transmission light have been magnified when the incident light
frequency is in conductance band, which can be made light
magnifier (magnifying multiple $10^{7}$).(2) For the negative
incident, i.e., the refractive indexes decreasing along the
direction of incident light, the intensity of transmission light
have been weaken, which can be made light attenuator (magnifying
multiple less than $1$). (3) The GFPCs structure (positive
incident FIG. 10 and negative incident FIG. 11) is the optical
diode, since the light intensity is magnified (the light get
across) when the light is positive incident, and the light
intensity has been decreased (the light cutoff) when the light is
negative incident, which transmits light from an input to an
output, but not in reverse direction.

\vskip 5pt

{\bf 6. Conclusion}

\vskip 8pt

In summary, We have theoretically investigated a new general
function photonic crystals (GFPCs), which refractive index is an
arbitrary function of space position. We choose the line
refractive index function for two mediums $A$ and $B$. By the
calculation, We obtain the following results: (1) when the line
function of refractive indexes is up, the transmissivity can be
far larger than $1$. (2) when the line function of refractive
indexes is down, the transmissivity can be far smaller than $1$.
(3) when the two-endpoint values of refractive index are equal for
mediums $A$ and $B$, the transmissivity $T$ is in the range of $0$
and $1$ (it becomes conventional PCs ). So, the conventional PCs
is the special case of the GFPCs. (4) when the refractive indexes
function increase along the direction of incident light, the light
intensity should be magnified, which can be made light magnifier.
(5) when the refractive indexes function decrease along the
direction of incident light, the light intensity should be weaken,
which can be made light attenuator. (5) The GFPCs can be made
light diode. The general function photonic crystals can be applied
to design more optical instruments.
\\

\end{document}